# Sentiment Progression based Searching and Indexing of Literary Textual Artefacts

Hrishikesh Kulkarni[1] and Bradly Alicea[2]

[1] SPPU Pune University, Pune 411016, India
[2] Bradly Alicea, Orthogonal Research & Educational Laboratory, USA
`hrishikeshparag@ieee.org`

**Abstract.** Literary artefacts are generally indexed and searched based on titles, meta data and keywords over the years. This searching and indexing works well when user/reader already knows about that particular creative textual artefact or document. This indexing and search hardly takes into account interest and emotional makeup of readers and its mapping to books. When a person is looking for a literary textual artefact, he/she might be looking for not only information but also to seek the joy of reading. In case of literary artefacts, progression of emotions across the key events could prove to be the key for indexing and searching. In this paper, we establish clusters among literary artefacts based on computational relationships among sentiment progressions using intelligent text analysis. We have created a database of 1076 English titles + 20 Marathi titles and also used database http://www.cs.cmu.edu/~dbamman/booksummaries.html with 16559 titles and their summaries. We have proposed Sentiment Progression based Indexing for searching and recommending books. This can be used to create personalized clusters of book titles of interest to readers. The analysis clearly suggests better searching and indexing when we are targeting book lovers looking for a particular type of book or creative artefact. This indexing and searching can find many real-life applications for recommending books.

**Keywords:** Literature, Creative Artifacts, Searching, Natural Language Processing, Text Analysis, Machine Learning, Information Retrieval, Sentiment Mining

## 1 Introduction and Related Work

### 1.1 Searching and Indexing literature

Searching, recommending and indexing of literary artefacts is generally driven by names of authors, topics and keywords. This is very effective but very primitive method and cannot cope up with uncertainty and variations associated with user interests. It comes with its own advantages and challenges. But when we take into account millions of unknown titles; does this indexing in true sense gives us book titles best suited to our interest and emotional makeup? Actually, these primitive indexing mechanisms create bias while making some titles best seller and do injustice to many classic literary creations by unknown champions. This even refrains newcomers from creating novel literary experiments. This makes many worthy literary creations to even sometime vanish behind the curtains of brands created by this name and author-based systems. While digital libraries are taking care of availability of books, we here propose an algorithm for fair indexing and recommending literary artefacts. The joy and satisfaction of reading has more to do with plot, theme, progression of emotional upheavals than title or name of author. Indexing based on sentiment progression, thematic changes could help in dealing with bias of indexing and making justice to many titles those could not reach to book lovers. Though titles may sometimes lead to selection, but in practical sense it has very little to do with reader's interest. In the ocean of books where more than two million books are published every year, many worthy titles fail to get even sight of relevant and interested readers. Digital media and intelligent indexing with sentiment progression-based recommendation can overcome this barrier. Availability, reachability and personalized indexing can help in taking a few steps towards solving this problem. This will reduce the bias, nurture creativity and bust the monopolies in the literary world.

### 1.2 Related Work

Indexing, listing and searching of books help readers in selecting the title of their interest. When we were focused on physical books – book catalogues are maintained alphabetically. These catalogues in digital platform were very useful for searching and locating a book if the title is known to a reader. These catalogues were extended with same paradigm of indexing for digital books. You can search even contents in these digital artefacts. But this basic paradigm of catalogues, indexing and searching has many limitations. You need to know a book if you are searching one. Keyword based search works effectively for scientific books but for fictions and narrative it fails miserably. The concepts of decoding character relationships for indexing and recommending is the core idea proposed in this paper. There are many attempts to decode relationships among characters. Decoding relationships among characters in narratives [1] can be considered as one of the core aspects while analyzing it. Relationship among characters at various places in a narrative is indicative of sentiment progression. These relationships can be modelled in different ways using semi-supervised machine learning [1]. Narrative structure is core to analyzing it [2]. The role of sentiment in this pattern mapping is crucial to such relationship models [3]. Topic transition is generally determined using consistency analysis and coherence [4].

Linguistic perspective and contextual event analysis can play a vital role in narrative assessment. Surprises bring unexpected changes in relationships and event progressions, differentiating adorable events [5]. The overall narrative can be viewed as an emotional journey with variations in interestedness. This journey progresses as various relationships in the given narrative unfold. In concept journey, different concepts are battling for existence and key concept proved to be ultimate winner. Emotional aspects blended in very personalized culture are at the helm of this journey. These emotional aspects are associated with part of stories or creative textual artefacts depicted through different impacting sentences [6]. While dealing with multiple stories, sentence similarity and word mapping can lead to initiation of analysis [7]. At later stage we need to deal with sentiment composition and progression from linguistic point of view [8]. Perspectives and context could prove to be of utmost importance in sentiment analysis [9].

Creative artefacts irrespective of genre need to be analyzed from different perspectives [10]. Here we typically need to analyze spoken language in social context, referred as discourse analysis. It deals with emotional exchanges during conversations among actors [11]. Different methods can be used for computational discourse analysis [12]. In this analysis dispute detection, competitive behavior of characters can be used as primary indicators. Dispute can go beyond mere conversation analysis [13]. Thus, text-based conversation or exchange of words can be analyzed [14]. In any of such scenarios decoding personality and cultural analysis with personality vector analysis prove to be effective for mapping [15, 16]. Researchers also used text-based analysis for clustering books [17].

The progression of relationships among characters in a narrative can be used for searching and indexing of books. This can take book catalogues beyond traditional limitation and hence searching can be possible based on progression of emotions, interestedness of readers and behavioral patterns of core characters in different titles. To make it possible, this paper focuses on core character identification and pivot point determination. The emotional progressions are classified and the representative types of progressions are used for indexing and searching. These progressions are selected based on the books loved by readers.

At times, book titles do not give information about the book and hence, indexing based on titles is useful only for them who already knows the title. But when a reader simply has a few titles of interest and does not know a specific title, it becomes difficult to search and locate a most compelling book of his/her interest. In such a scenario how one can go and catch the most relevant book out of millions of published books. This paper proposes 'Sentiment Progression based Search and Indexing' (SPbSI) model to solve this problem. Themes, sentiments and culturally relevant touchstones form the basis for any story. There could be strong and weak emotions in the story and as the book progresses the sentiment progression takes place with emotional revelations. Thus, the books are clustered based on sentiment progression. They are further indexed based on the representative sentiment progression and distance of every book in the cluster from the representative progression.

## 2 Sentiment Progression based Search and Indexing Model (SPbSI)

The proposed method is divided into four important phases:
1. Primary keyword-based core character identification and selection of pivot points
2. Sentiment progression analysis across pivot points
3. Derive similarity using 'Sentiment Progression Similarity Indicator' for clustering of sentiment progressions.
4. Indexing and preparing catalogue for effective sentiment progression search

## 3  Data Analysis

A database of (1076 English + 20 Marathi) book titles from different genre is prepared and used for testing and learning. The book list can be accessed from location: https://drive.google.com/file/d/1uWjRWBYfOZhH2XpFfO-VONYf_L3l1uw0/view. Another database used is http://www.cs.cmu.edu/~dbamman/booksummaries.html. One of the sample books we used as an example - titled "*Rage of Angels*" is a part of both of these datasets. The analysis on this data is preformed using SPbSI to determine relevance of indexed items and personalized search results for book lovers.

## 4  Mathematical Model

### 4.1  Core character identification and Pivot Point Selection:

Core characters are the characters those are crucial to narrative and the story cannot progress without them. They directly or indirectly participate in all important events across the narrative. They are identified based on their frequency and relationships with other characters. They are generally very well connected to main theme and different key events across the narrative.

To explain the concept, we have chosen two interesting fictions those were read by 50 out of 150+ book lovers from the BDB book club[1]: First one is the fiction published in 1980 titled '*Rage of Angels*' (https://en.wikipedia.org/wiki/Rage_of_Angels) by Sidney Sheldon & the second one is Marathi Classic *KraunchVadh* by Gyanpeeth awardee writer V.S. Khandekar (https://en.wikipedia.org/wiki/Vishnu_Sakharam_Khandekar).

The algorithm to identify core characters is built around core words and pivot points. Here core word is defined as a word that belongs to keyword set and has highest frequency of occurrence across the text space of interest. This word acts as a reference while creating cluster of words. Similarly, Core Character (CC) is one of the prime characters in narrative and is defined based on its presence and association with other prime characters. When more prime characters get associated with the character of interest, the weight associated with it increases.

Equation 1 gives mathematical definition of CC.

$$\forall c \in c | C \in [CC] \text{ and } c \to [CC] \text{ where } [CC] \neq \Phi \qquad (1)$$

Going through narrative in an iterative fashion, the core characters are identified. The characters Jenifer, Michael and Adam are identified as core characters in *Rage of Angels*, while *Kraunch-Vadh* has core characters Sulu, Dilip, and Bhagvantrao. Pivot point is a location in a narrative where we perform sentiment analysis. It is a place marked by presence of intense interactions among two or more core characters.

---

[1] BDB Book Club is a major book club run by BDB India Pvt Ltd in Pune https://bdbipl.com/index.php/bdb-book-club/

The text or book is divided into logical blocks where logical blocks are delimited by pre-decided number of tokens. Let's continue with our example KraunchVadh. To further elaborate approach, with coreference resolution we find the occurrence weight of 27 at pivot point at logical block 62 occurring on page 57-58 where Dilip meets Sulu. Its peak is at the exclamation of Dilip "I had attraction outside to come out of Jail – one of my mother and other of…" he points out to beautiful image of Sulu in the mirror.

Similarly, we can find occurrence weight of 24 at pivot point at logical block 92 on Page 87 for relationship between Bhagvantrao and Sulu. Here weight represents normalized keyword occurrence in the proximity of pivot point.

### 4.2 Sentiment Progression Analysis across Pivot Points

The sentiment and emotional index at a pivot point using expressive word distribution is used to derive sentiment index. Thus, analysis of relationships helps to derive overall sentiment index. The progression of sentiment across these pivot points represents the behavior and nature of narrative.

**Algorithm 1.** Indexing for effective sentiment progression search

Task 1: Initialization (Determine Pivot Points)
1.    $CC_{NAR} = \forall c \in c | C \in [CC]$ and $c \to [CC]$ where $[CC] \neq \Phi$

Task 2: Identify Pivot Point based on frequency, presence and Sentiment
2.    $for\ \forall SJ_i,\ CC_i\ where\ CC\ ] \neq \Phi\ \&\&\ j \leq PP\_max\ \ do$
3.       $SI_i[SJ_i CC_i] = Sentiment(CC_i) + \alpha$
4.       $With\ gradient\ decent\ determine\ local\ sentiment\ maxima$
5.       $PP_j = Local\ Maxima$
6.       $j++$
7.    $end$

Task 3: Determine Sentiment Progression (based on Sentiment Value at Pivot Point (PP))
8.    $for\ \forall PP\ do$
7.       $if([SV@PP] \neq \Phi)$
8.          $Insert\ SV\ in\ to\ series$

Task 4: Cluster formation of Sentiment Progression
9.    $Sentiment\ Distanc\ SD = \frac{\sum_{i=1}^{n} CF(i)^2 \times N(i)}{\sum_{i=1}^{n} N(i)}$
10.   $Sentiment\ Progression\ Similarity\ Indicator\ SPSI = \frac{1}{(1+\ln(1+SD))}$
11.   $for\ \forall Series\ IF\ [(Series_{Length} <> M)]$
12.       $Make\ length\ of\ series = M$
13.   $for\ \forall Series\ where\ [Series] \neq \Phi\ \&\&\ Series\ length = M$
14.       $Create\ SPSI\ Matrix$
15.       $for\ \forall Series\ where\ [Series] \neq \Phi\ \&\&\ \exists(SPSI) > DT$
16.          $IF\ [(SPSI(i,j) > Dynamic\ Threshold\ DT)]$
17.             $Combine\ Seires\ (i,J)$
19.       $endif$
20.   $end$

End of Algorithm

Pivot point detection algorithm has identified 10 pivot points across the novel for characters Sulu and Dilip. Similarly, there are 8 pivot points for Sulu and Bhagwantrao. The progression of sentiment across these pivot points is determined. The progression of relationship between Dilip and Sulu is predominant one. The detail algorithm SPbSI for indexing based on pivot point determination and sentiment association is given in algorithm 1. The sentiment progressions take into account sentiment and emotion indicators at pivot points. Emotional upheavals involving the core characters contribute majorly to this pattern.

The extraction of sentiment (emotional positivity and negativity in this case) with reference to context of story includes pivot point identification, extracting sentiment at a particular event. These sentiments are progressed from one pivot point to next one. This progression is represented on a timeline as a pattern. These patterns are input for clustering and deriving the representative pattern.

This is determined at different pivot points also. Thus, Model' (SPbSI) identifies sentiment progression from one pivot point to another.

### 4.3 Derive Similarity Using Sentiment Progression Similarity Indicator

Statistically Sentiment Progression Similarity Indicator (SPSI) gives behavioral similarity between two sentiment progression patterns. Every pivot point has a sentiment value. Hence every book has a sentiment progression series and can be represented as a data series. This makes it possible to represent every book as a sentiment progression series. Let's take two creative textual artefacts at a time and get two sentiment progression series corresponding to these two artifacts. It is highly likely that these two series will have different number of pivot points. Hence, we add supporting points so that both series have equal number of elements. Hence series will look like:

$$RS \ni S(i) = S_1(i) + S_2(i) \tag{5}$$

Here, RS is a series derived by summing corresponding pivot point sentiment values. This is used to calculate the probable value for corresponding series. Thus, probable sentiment value in accordance with sentiment progression is determined using Eq 6.

$$PS = \frac{\sum_{i=1}^{n} S_1(i)}{\sum_{i=1}^{n} S(i)} \tag{6}$$

PS is used to derive expected sentiment value with assumption that sentiment progression in second series is same. It us used to calculate correction factor CF.

$$CF(i) = \frac{PS \times RS(i) - S_1(i)}{\sqrt{RS(i) \times PS \times (1 - PS)}} \tag{7}$$

The sentiment distance SD between two text artefacts is given by Eq 8.

$$SD = \frac{\sum_{i=1}^{n} CF(i)^2 \times N(i)}{\sum_{i=1}^{n} N(i)} \tag{8}$$

Here N is normalization factor $N(i) = \sqrt{R(i)}$.

Further, the Sentiment Progression Similarity Indicator (SPSI) is calculated using Eq 9

$$SPSI = \frac{1}{(1+\ln(1+SD))} \quad (9)$$

It makes sure that 'Sentiment Progression Similarity Indicator' (*SPSI*) will drop slowly with increase in sentiment distance. Value of *SPSI* is close to 1 for patterns those look alike while it approaches to zero for patterns those are completely different. The SPSI with self-pattern or for exactly identical patterns is 1.

### 4.4 Indexing and Preparing Catalogue for Effective Sentiment Progression Search

Iteratively similarity between sentiment progression of every pair of creative textual artifacts is calculated. This leads to SPSI matrix. The diagonal of this matrix is always 1. Two series with maximum similarity are combined to reduce the (n X n) matrix to (n-1 X n-1) and so on. This process continues till the similarity between all representative patterns is less than dynamic threshold. This process results in getting clusters with representative sentiment progression patterns. A Progression Similarity Matrix for nine books is depicted in Fig 1.

Here out of these 9 series during first iteration series (7, 9) are combined. Further the resultant series is combined with series 3, then with series 6 and later with series 5. Thus a representative series is formed for cluster made up of series (3, 5, 6, 7, 9). Similarly, series (1, 4) are combined and that series is combined with series 8. Thus, a cluster is formed of series (1, 4, 8). Thus, at the end of iteration 1 the matrix will be of size 3 X 3, with members representing cluster (3, 5, 6, 7, 9), cluster (1, 4, 8) and (2).

| Series | 1 | 2 | 3 | 4 | 5 | 6 | 7 | 8 | 9 |
|---|---|---|---|---|---|---|---|---|---|
| 1 | 1 | 0.32 | 0.11 | 0.76 | 0.36 | 0.16 | 0.15 | 0.62 | 0.11 |
| 2 | 0.32 | 1 | 0.18 | 0.22 | 0.31 | 0.28 | 0.14 | 1 | 0.23 |
| 3 | 0.11 | 0.18 | 1 | 0.16 | 0.58 | 0.54 | 0.73 | 0.25 | 0.41 |
| 4 | 0.76 | 0.22 | 0.16 | 1 | 0.37 | 0.26 | 0.39 | 0.57 | 0.25 |
| 5 | 0.36 | 0.31 | 0.58 | 0.37 | 1 | 0.49 | 0.52 | 0.16 | 0.53 |
| 6 | 0.16 | 0.28 | 0.54 | 0.26 | 0.49 | 1 | 0.66 | 0.11 | 0.50 |
| 7 | 0.15 | 0.32 | 0.73 | 0.39 | 0.52 | 0.66 | 1 | 0.29 | 0.86 |
| 8 | 0.62 | 0.14 | 0.25 | 0.57 | 0.16 | 0.11 | 0.29 | 1 | 0.15 |
| 9 | 0.11 | 0.23 | 0.41 | 0.25 | 0.53 | 0.50 | 0.86 | 0.15 | 1 |

Figure 1: Progression Similarity Matrix.

Thus, each representative cluster pattern is converted in to an index point. The all narratives belonging to that cluster are represented graphically close to that index point. The similar ones are represented closer to it and organized based on distance from representative behavior.

### 4.5 Handling Unequal Length Pivot Point Data Sets

Handling unequal length data series is one of the most challenging aspect of this method. To deal with this, we distributed pivot points based on its distribution across the book for the shorter length data series. The biggest gap is filled with intrapolation first. This process is continued till the length of two data series becomes same. The 30% length difference can be handled with this method but the method fails for more than 30% length difference between two series. In this case the secondary pivot points are used while extrapolating. A secondary pivot point between two primary pivot points is selected and the values are adjusted with reference to primary adjacent pivot points.

## 5 Experimentation

### 5.1 Baselines

Creative artifacts are generally indexed and catalogued using titles or author names. In some very special cases support is provided using metadata and keywords. Thus, indexing in the past performed using two different approaches [17]. In the first approach it is based on metadata, author names, genre, keywords and titles. In the second approach textual similarity is used across the complete text or on the summary of the books for clustering. The first approach is developed as the baseline-1 while the second one is developed as baseline-2. The results of these baselines are compared with SPbSI to validate suitability of the proposed approach.

### 5.2 Results

Table 1 gives sentiment indices at normalized pivot points for fictions *Rage of Angels* and *KraunchVadh*. The Sentiment Progression Similarity Indicator (SPSI) between these two fictions is 0.649149. Fig. 2 depicts sentiment progression for them across pivot points.

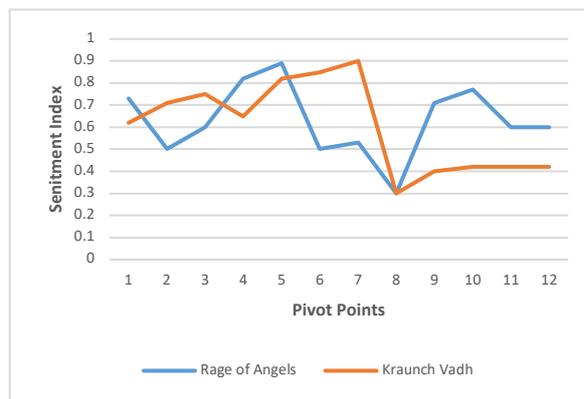

**Fig. 2.** Comparison of Sentiment progression

**Table 1.** Pivot Point Mapping and Ranking across the fiction (Normalized values)

| Pivot Points | Sentiment Index: *RageofAngels* | Sentiment Index: *KraunchVadh* |
| --- | --- | --- |
| 1 | 0.73 | 0.62 |
| 2 | 0.5 | 0.71 |
| 3 | 0.6 | 0.75 |
| 4 | 0.82 | 0.65 |
| 5 | 0.89 | 0.82 |
| 6 | 0.5 | 0.85 |
| 7 | 0.53 | 0.9 |
| 8 | 0.3 | 0.3 |
| 9 | 0.71 | 0.4 |
| 10 | 0.77 | 0.42 |
| 11 | 0.6 | 0.42 |
| 12 | 0.6 | 0.42 |

Books from database (http://www.cs.cmu.edu/~dbamman/booksummaries.html) are used for experimentation. We needed complete text of the book, hence the number of samples used for experimentation are kept limited. The response of 25 book lovers is compiled for analysis of outcome. Total 100 top books are indexed and catalogued using SPbSI. This outcome is compared with results from baseline algorithm where book lovers look for a book of a particular type. These books are clustered in Three types of behavioral patterns. The representative behavioral patterns are depicted in Table 2.

100 books are used for clustering and Indexing. This indexing based on clusters is verified using inputs from book lovers. Out of these 100 books for 92 all the book lovers were in agreement indexing and mapping. On the other side baseline-1 algorithm based on TFIDF using author names and metadata found only 55 booklovers endorsed the outcome. Baseline-2 uses text similarity algorithm based on book summaries which could find 66 books indexed as per expectations of the book lovers. The behavioral pattern and indexing are depicted in Table 2. Around 38% improvement could be obtained for the given set of data using SPbSI over the baseline-1 and 26% over baseline-2.

This sample response endorses that readers look for sentiment progression rather than metadata related to book. Which definitely creates an avenue for searching and indexing books based on behavioral sentiment patterns rather than titles or author names. In another enhancement the title and author name can be used in conjunction with primary key of behavioral sentiment progression. Though the sample size is small one, it is representative of overall similarity.

**Table 2.** Sentiment Progression based Indexing

| Sr. No. | Representative Sentiment Progression Behavioral Pattern | Example Books | Remarks |
| --- | --- | --- | --- |
| 1 | *[line chart, x: 1–15, y: 0–100]* | *Rage of Angels, KraunchVadh*, etc | Middle portion creates higher sustained positive emotions with peak at pivot point 11 |
| 2 | *[line chart, x: 1–15, y: 0–100]* | *Nothing lasts forever, Amrutvel*, etc | Slowly leads to lower point and there are surprises and excitement towards end |
| 3 | *[line chart, x: 1–15, y: 0–80]* | *Alchemist, Rikama Devhara*, etc. | Distributed surprises excitement leading to multiple spikes across the narrative |
| 4 | *[line chart, x: 1–15, y: 0–100]* | *Kite Runner, Five Point Someone*, etc. | Begins with excitements and multiple spikes and followed by sudden negativity with some positive resolve towards end |

## 6   Conclusion

Indexing and searching narratives and creative textual artefacts using author names and titles comes with its own challenges. It has been very useful when we are looking for a particular title or book. But someone is looking for a behavior of a narrative without knowledge of a particular title he/she may not able to get it with this traditional indexing mechanism. In such a scenario sentiment progression could prove to be a valid alternative to this traditional way of indexing or searching of books. Book similarity in terms of reader preferences depends on progression of sentiment. This paper proposed an approach of indexing and searching of books based on 'Sentiment Progression based Searching and Indexing' (SPbSI). The results are analyzed with reference to data collected from 25 book lovers, but the method may be scaled to the analysis of thousands of candidates. The proposed algorithm gives around 26% improvement over the base line algorithm. The algorithm SPbSI can further be improved using moving window-based similarity approach for the next level of sentiment progression-based similarity detection. Which can make possible even to recommend certain part of a narrative or creative artifact to readers. This will definitely demand for a greater number of pivot points or region-based pivot points. The promising results definitely give path to new experimentation opportunities. This method can find its applications in searching books based on interest or with reference to a particular title. It can even evolve to application that can help in building recommendation systems based on multiple perspectives. Extending this idea for movies and plays can lead to new direction in creating movie and play catalogues. This can further lead to possible research in creating catalogues across different types of creative artifacts with more systemic value.